\documentclass[twocolumn]{aastex631}

\usepackage{blindtext}
\usepackage{amsmath,amssymb}

\begin{document}

\title{Far-Ultraviolet Emission Line Investigation of Flares on AU Mic}

\author[0009-0006-5310-6855]{Adalyn Gibson}
\affiliation{Department of Astrophysical and Planetary Sciences, University of Colorado, UCB 389, Boulder, CO 80309, USA}
\affiliation{Laboratory for Atmospheric and Space Physics, University of Colorado Boulder, UCB 600, Boulder, CO 80309}

\author[0000-0001-7458-1176]{Adam F. Kowalski}
\affiliation{Department of Astrophysical and Planetary Sciences, University of Colorado, UCB 389, Boulder, CO 80309, USA}
\affiliation{Laboratory for Atmospheric and Space Physics, University of Colorado Boulder, UCB 600, Boulder, CO 80309}
\affiliation{National Solar Observatory, University of Colorado Boulder, 3665 Discovery Drive, Boulder CO 80303, USA}

\author[0000-0002-9464-8101]{Adina~D.~Feinstein}
\altaffiliation{NHFP Sagan Fellow}
\affiliation{Department of Physics and Astronomy, Michigan State University, 567 Wilson Rd, East Lansing, MI 48824}

\correspondingauthor{Adalyn Gibson;} \email{adgi9680@colorado.edu}

\begin{abstract}
The role of non-thermal proton energy transportation during solar and stellar flares is largely unknown; a better understanding of this physical process will allow us to rectify longstanding deficiencies in flare models.  One way to detect the presence of non-thermal protons during flares is through the Orrall-Zirker (OZ) effect, proposed by \cite{oz1976}, whereby an enhanced red wing appears in hydrogen emission lines (e.g., Lyman-$\alpha$ at 1215.67 \AA).  We analyze archival \textit{Hubble Space Telescope}/Cosmic Origins Spectrograph G130M (1060 - 1360 \AA) observations of the young M dwarf, AU Mic to search for evidence of OZ effect during the impulsive phase of six stellar flares with $E_\textrm{flare} \approx 10^{30 - 31}$\ erg. While we found non-detections of the OZ effect, we note there is a pronounced blue enhancement in several \ion{C}{2} and \ion{C}{3} emission lines during one of the high-energy flares.  We propose that either filament eruptions or chromospheric evaporation could be the mechanism driving this observed blue enhancement. We compare the far-ultraviolet (FUV) spectra to 1D radiative-hydrodynamic stellar flare models, which are unable to reproduce the blue enhancement and broadening in these cool flare lines. By completing a line-by-line analysis of the FUV spectrum of AU Mic, we provide further constraints on the physical mechanisms producing stellar flares on M dwarfs.

\end{abstract}


\section{Introduction}\label{sec:intro}

Our knowledge of flare-accelerated particles during magnetic reconnection events on other stars has significant gaps. The mechanism behind particle acceleration in flares, the resulting particle energy distributions, and in particular the role of non-thermal protons are poorly understood and are lacking in current flare models \citep{Kowalski_review}. Proton beams may constitute up to or exceeding half of the energy during particle acceleration in solar flares \citep{2012Emslie}. If this is the case, then our current flare models are missing an important constituent of the energy transport through the solar/stellar atmosphere \citep{kerr23}. Non-thermal protons are thought to deposit their energy deeper in the atmosphere than non-thermal electrons, thus possibly providing an explanation for the enhancements in the optical continuum, commonly referred to as white-light, that have a photospheric origin  \citep{2024Sadykov, 2014kerr, 2013Watanabe, 1989Neidig, 2018Proch}. 

The origin of white light flares is still unknown \citep{Watanabe:2017}. Non-thermal electrons alone cannot account for a photospheric origin of white-light flares because they cannot penetrate deep enough into the atmosphere to cause the required heating \citep{2013Watanabe}. Chromospheric evaporation occurs when non-thermal electrons stream along magnetic field lines and impact the chromosphere. Once these non-thermal electrons arrive at the chromosphere, they deposit their energy into the plasma, heating the plasma and increasing pressure. The heated plasma is accelerated upwards, resulting in chromospheric evaporation \citep{1985Fisher, 1985Fisher02, 1985Fisher03}.

Observational evidence of chromospheric evaporation includes blue-shifted coronal spectral lines during the impulsive phase of the flare \citep{Milligan&Dennis, 1980Doschek, 1984Antonucci, shibata1995, 2015Graham&Cauzzi, 2016Polito} and the empirical relation between impulsive and gradual phase emissions known as the Neupert effect \citep{1968Neupert, 1993Dennis&Zarro, 2003Veronig}.  There are generally thought to be two regimes of chromospheric evaporation:  gentle and explosive, whereby the dividing line is a complicated function of the injected electron beam properties; \citep[flux, low-energy cutoff, and power-law index][]{1989Fisher, 2015Reep}.  Solar models of chromospheric evaporation tend to favor heating driven by electron beams rather than thermal conduction \citep{2016Polito}.

In stellar flares, blueshifts have also been observed in hot coronal flare lines \citep{2019Argiroffi, 2024Inoue}, but occasionally they also appear in cooler chromospheric lines such as H $\alpha$ \citep{2008Fuhrmeister, 2018Honda, 2024Notsu, 2025Lu, 2025Kajikiya}.  Blueshifts are typically modeled as filament eruptions or coronal mass ejections in stellar flares.  Radiative-hydrodynamic models of chromospheric evaporation have been able to produce blueshifts in cool lines \citep{2005Allred}, but the evaporations are too short lived and too slow to explain most of the stellar data. More recent radiative-hydrodynamic models suggest that changes in the electron beam can drive deeper material from the chromosphere into the transition region and corona \citep{2024kerr&polito, 2023Polito}.

To determine the cause of white-light flares and other responses to deep energy deposition, we seek also an accurate picture of the non-thermal proton acceleration process. Constraining the presence of low-energy protons among the accelerated particle population in turn constrains the total energies available to explain the white-light radiation and other flare emission.  An empirical re-examination of the evidence for these low-energy protons is timely because there has been significant  progress in producing power-law spectra of protons extending down to low energies during reconnection \citep{2024Yin} and in modeling the warm-target transport of low-energy protons from the corona to the chromosphere \citep{2020Allred}. Furthermore, re-examination of gamma-ray images of large solar flares now suggests the sources of non-thermal electron and proton impact with the chromosphere to be cospatial \citep{2025Battaglia}, thus transforming decades of understanding \citep{2016Hudson}.

One possible observational signature of non-thermal protons accelerated during stellar flare events is the Orrall-Zirker (OZ) effect \citep{oz1976}. The OZ effect occurs when non-thermal protons are accelerated downwards in the corona to the mid-upper chromosphere. In the chromosphere, the non-thermal protons charge exchange with ambient hydrogen atoms, thus producing a population of non-thermal energetic neutral atoms (ENAs). These ENAs subsequently go through spontaneous radiative de-excitation and emit Doppler redshifted photons in the red wings of hydrogen spectral lines \citep[Figure 4 in ][]{kerr23}. Tentative evidence of the OZ effect has been seen in a single star --- AU Microscopii \citep[AU Mic;][]{woodgate1992}. AU Mic is a $22 \pm3$ Myr \citep{age} M-dwarf \citep[M1Ve; R$_\star/$R$_\odot = 0.862 \pm 0.052$; M$_\star/$M$_\odot = 0.50 \pm 0.03$][]{spectype, 2020Nature, 2022_new}. AU Mic is known for its enhanced stellar activity due to its young age, including a high flaring rate \citep{feinstein22, gilbert22, 2023Tristan}, short rotation period \citep[$4.863 \pm 0.010$ days;][]{2020Nature}, and strong magnetic field \citep{donati23}. AU Mic has been the target for several flare-driven observational campaigns \citep{2023Tristan, paudel24, 2025Tristan} and is a standard for modeling M-dwarf flares due to its close proximity \citep[$9.714 \pm 0.002$ pc; ][]{gaia_2021} and high signal-to-noise observations. 

\cite{oz1976} predicted a signature of non-thermal proton beams accelerated during flares could be observed as an increase in the red wing of Lyman-$\alpha$ (Ly$\alpha$; 1215.67 \AA). \cite{kerr23} defined the expected wavelength range to detect the OZ effect as  5.6 - 17.7 \AA\ redward (1380-4365 km s$^{-1}$) of Ly$\alpha$ \citep[see also][]{1985Canfield}. The OZ effect would be detectable for low energy non-thermal protons with energies ranging from $\sim$ 0.01 to 1 MeV as discussed in \cite{kerr23} and \cite{1985Canfield}. \cite{woodgate1992} presented a tentative detection of the OZ effect on AU~Mic through an observed red enhancement in Ly$\alpha$ using the Goddard High Resolution Spectrograph on the \textit{Hubble Space Telescope} (HST).  The detection lasted for 3~s at the flare start, which is in agreement with the requirement for a relatively large area of undisturbed atmosphere. To date, this is still the only tentative detection of the OZ effect on another star \citep{kerr23, 2001Robinson}. This tentative detection highlights that HST could detect the OZ effect here if a sufficient amount of non-thermal protons are accelerated during a flare ($10^9 - 10^{11}$ erg s$^{-1}$ cm$^{-2}$, as modeled in \cite{kerr23}). \cite{feinstein22}  performed a preliminary search in recent HST/Cosmic Origins Spectrograph (COS) observations and did not detect the presence of the OZ effect. 

In addition to its activity, AU Mic has been a target of observational campaigns due to the presence of two transiting exoplanets \citep{2020Nature, 2021_martioli} and its debris disk \citep{liu2004, lawson23}. AU Mic b is a Neptune-like exoplanet (R$_p = 4.20 \pm 0.20$ R$_\oplus$) on a close orbit, with a semi-major axis a$ = 0.0645 \pm 0.0013$~au. AU Mic c is slightly smaller (R$_p = 3.24 \pm 0.16$ R$_\Earth$) and farther than AU Mic b, with a semi-major axis of $0.1101 \pm 0.0022$~au \citep{2021_martioli}. Due to the close proximity of both planets, their evolution has the potential to be shaped by AU Mic's frequent stellar flares \citep[e.g.,][]{feinstein22, doAmaral2025}. For example, M-dwarf stars' frequent and intense flaring may be a source of atmospheric stripping of exoplanets \citep{lammer07}, with exoplanet atmospheric loss being predicted to be partially driven by M-dwarf flaring activity \citep{2022Neves_Ribeiro}. The potential of later atmospheric recovery is debated \citep{2019Tilley}. On the contrary, frequent flaring may jump-start the formation of organic molecules and prebiotic chemistry necessary for the evolution of life \citep{2024dransfield}; simulations of superflares on the young Sun suggest they may have been instrumental in forming the molecules essential for life \citep{2016Airapetian}.

Beyond the AU~Mic system, a complete understanding of the underlying physics driving magnetic reconnection events and the resulting observable stellar flares will provide a comprehensive understanding of the environments in which most planets evolve. M-dwarf stars such as AU Mic are known for their frequent and intense flaring throughout and especially early in their lives \citep{feinstein2020, 2024crowley}. Stellar CMEs, often accompanied by flares, can potentially be associated with spectral line shifts during a stellar flare, as shown in \cite{ikutash2024}. Due to the close-in habitable zones around M-dwarfs, atmospheric erosion of potentially habitable planets can be brought about by stellar CMEs from M-dwarfs more easily than CMEs from other spectral types of stars. It is crucial to understand the frequency and time evolution of stellar CMEs from M-dwarfs in order to determine their impact on atmospheric loss and evolution \citep{2007Khodachenko, lammer07}. The current models of stellar flares are not conclusive or complete enough to give accurate insight into the effects of stellar flares on exoplanet atmospheres. Bridging the current gaps in how flares operate will allow an understanding of exoplanet atmospheric responses. The models used in \cite{Chen:2021}, where the effects of flares on rocky planets are explored, only use photometric information, so there is no spectroscopic component to these models. By neglecting a spectroscopic component, the full impact of the effects of flares on rocky exoplanets are overlooked \citep{2025Kowalski}. We do not have many exoplanet spectra that show the direct response to flare irradiation to bridge this gap; thus the models of how flares affect exoplanet atmospheres do not leverage the precision that these spectroscopic observations offer.

In this work, we present a reanalysis of three HST/COS visits of AU~Mic \citep{feinstein22, feinstein24} to search for evidence of non-thermal proton beams. We present newly identified flares in the third visit. This work is laid out as follows. We describe the HST observations, data reduction, and emission line identification in Section~\ref{sec:obs}. We identify and characterize new stellar flares in Section~\ref{sec:flares}.
In Section~\ref{sec:SearchOZ}, we present the results of our search for evidence of the OZ effect and quantify an identified blue enhancement in several \ion{C}{2} and \ion{C}{3} emission lines. In Section~\ref{sec:discussion} we speculate further on the cause of the identified carbon enhancement and discuss the implications of this work on stellar flare models. We conclude in Section~\ref{sec:conclusion}.

\begin{figure*}[!htb]
    \centering
    \includegraphics[width=\textwidth]{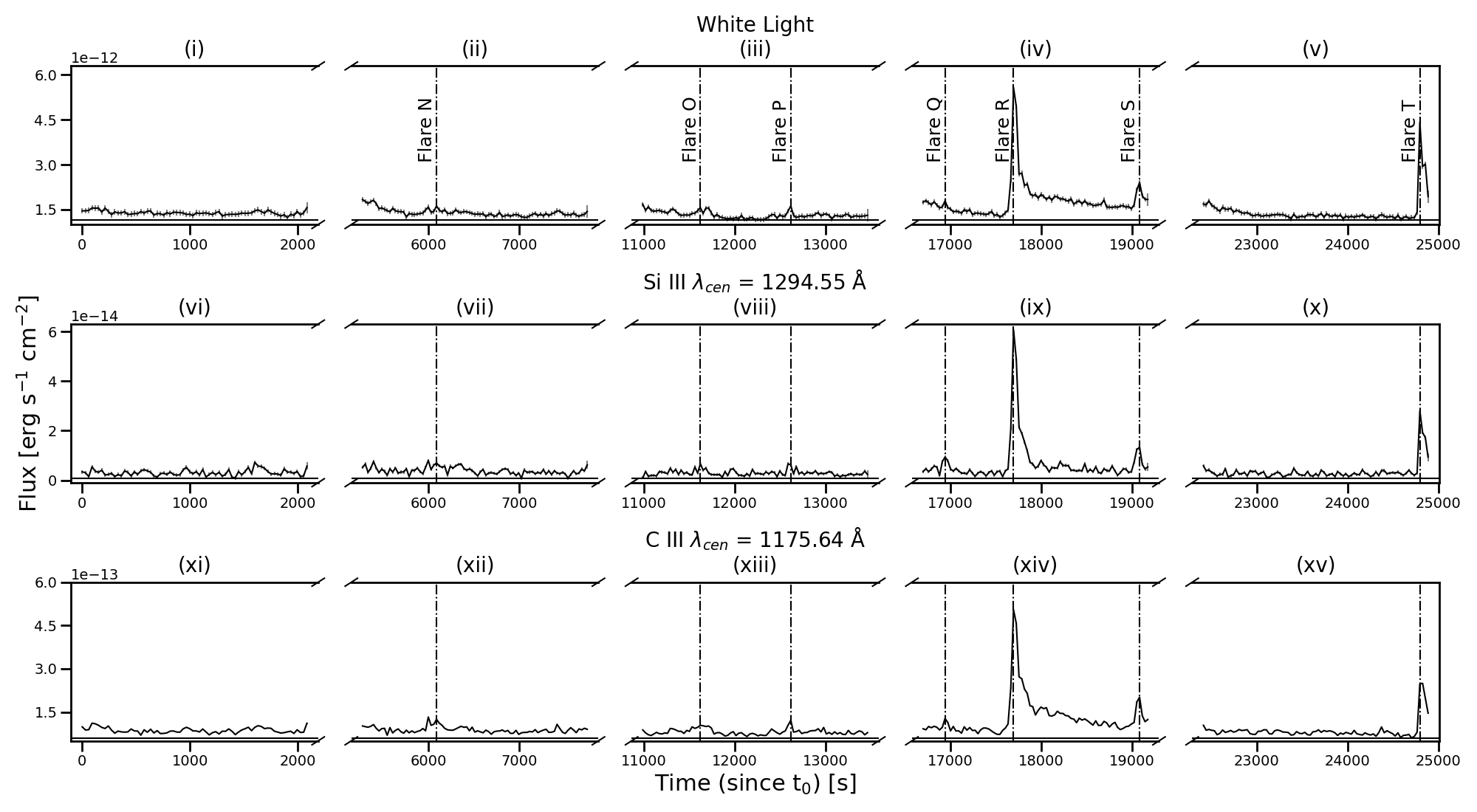}
    \caption{Identified flares from AU~Mic from Visit 3. We present three light curves per visit: white light (top), \ion{Si}{3} ($\lambda_{cen}$ = 1294.55 \AA; middle), and \ion{C}{3} ($\lambda_{cen}$ = 1175.64 \AA; bottom). Flares are labeled and marked by vertical dashed lines. We identified seven new flares in this visit. The new flare parameters can be found in Table~\ref{tab:Flares_vis3}. We note that even though these observations were taken a year after Visits 1 and 2, AU~Mic shows a similar flaring rate ($\sim 7$~flares visit$^{-1}$). The horizontal black line in each figure is the quiescent flux for each wavelength range. }
    \label{fig:visit3}
\end{figure*}

\section{Observations}\label{sec:obs}

We observed three visits of AU~Mic with HST/COS G130M ($\lambda_\textrm{cen} = 1222$\AA; $\lambda_\textrm{full} = 1065 - 1365$\AA, Resolving Power, R= $\lambda/\Delta \lambda_{\rm{FWHM}}= 12,000 -17,000$) under GO 16164 (PI Cauley). Visit 1 was executed on  2021-05-28 from 03:33:31 ---  15:50:43, Visit 2 was executed on 2021-09-23 from 15:06:14 --- 22:02:08, and Visit 3 was executed on 2022-10-09 from 12:38:55 ---  19:33:40. The instrument configuration for all three visits is identical. All visits have a total exposure duration of 2085 seconds for the first orbit and an exposure duration of 2471 seconds for the subsequent orbits. 

COS has a unique \texttt{time-tag} mode, which tracks the time at which each photon hits the detector. Through this mode, we are able to generate time-tagged spectra and light curves. We used the \texttt{\detokenize{costools.splittag}}\footnote{\url{https://github.com/spacetelescope/costools/tree/master/costools}} and  \texttt{\detokenize{CalCOS}}\footnote{\url{https://github.com/spacetelescope/calcos/tree/master/calcos}} pipelines to process the data into wavelength-calibrated spectra and analyzed the spectrum from each \texttt{corrtag} file, using these spectra to create spectroscopic light curves. We generated light curves at a 30-second cadence, which is sufficient for studying the morphology of flares. While we explored light curves at a higher cadence, the signal-to-noise (SNR) of individual emission lines became too low. Thus, the 30-second light curves provided a balance between a high time resolution and sufficient SNR to analyze  specific emission lines. We note that the wavelength calibration region for the fourth orbit for Visit 3 was affected by a hot spot, a transient region of high detector background. The standard \texttt{costools} and \texttt{CalCOS} pipelines were unable to fully reduce this visit while the hot spot was present. As such, in order to uniformly reduce this orbit with the others during Visit 3, we masked the hot spot in the \texttt{rawtag} file. This hot spot occupies a region of the detector between 6850 $<$ RAWX $<$ 6950 and 570 $<$ RAWY $<$ 610.

AU~Mic has a very high SNR FUV spectrum, allowing for the identification of over 100 individual emission lines \citep{feinstein22}. For the purposes of our investigation, we used several emission lines: \ion{C}{3} six-tuplet ($\lambda_{cen}=1175.64$~\AA), Ly $\alpha$, and \ion{C}{2} doublet ($\lambda_{cen} = 1335.095$~\AA). 

\section{The FUV Stellar Flares of AU~Mic}\label{sec:flares}

Understanding the time evolution, energy, and rate of flaring for AU Mic is important in understanding specific physical phenomena, such as the OZ effect, observed during flares. In this work, we present new stellar flares which occurred during Visit 3. We followed the identification technique described in \cite{feinstein22}. \cite{feinstein22} identified flares in Visits 1 and 2 by identifying outliers in the \ion{Si}{3} ($\lambda_{cen}$ = 1294.55 \AA) and \ion{C}{3} ($\lambda_{cen}$ = 1175.64 \AA) emission line light curves, presented in Figure~\ref{fig:visit3}. Flares were identified by their well-known time-dependent behavior of a sharp initial increase followed by an exponential decay phase. We identify an additional seven flares in Visit 3; we follow the same naming convention and introduce flares N - T. 

We uniformly calculated the energies and equivalent durations for flares A - T, where flares A - M occurred in visits 1 and 2. We derive the flare energy (E) using the following prescription:

\begin{equation}\label{eqn:1}
    E = 4 \pi d^2  \int_{t_0}^{t_1} F_f(\tau) - F_q(\tau) \,d\tau    \quad   (1)
\end{equation}

where $d$ is the distance to the star, $t_0$ is the start time of the flare defined as the time point the flux increases by 1-$\sigma$ from the quiescent flux, $t_1$ is the stop time of the flare defined as the time point the flux gets within 1-$\sigma$ of the average quiescent flux, $F_q$, and $F_f(\tau)$ is the total flux during the defined start and stop times. The 1-$\sigma$ fluxes above $F_q$, are defined as $F_q$+$7.94^{-14}$\ erg\ s$^{-1}$\ cm$^{-2}$ for the white-light flux, $F_q$+$6.20^{-16}$\ erg\ s$^{-1}$\ cm$^{-2}$ for the \ion{Si}{3} flux, and $F_q$+$1.02^{-15}$\ erg\ s$^{-1}$\ cm$^{-2}$ for the \ion{C}{3} flux. We adopted $d = 9.714 \pm$ 0.002\ pc \citep{gaia_2021}. To calculate the equivalent duration (ED), first described by \cite{1972Gershberg}, we used the following prescription:

\begin{equation}\label{eqn:2}
    ED = \int_{t_0}^{t_1} \frac{F_f(\tau) - F_q(\tau)}{F_q(\tau)} \,d\tau    \quad   (2)
\end{equation}

We calculated the energies and durations of the new flares in the ``white light" (broadband) wavelength-integrated flux ($\lambda = [1080, 1360]$~\AA), \ion{Si}{3} wavelength-integrated line flux ($\lambda = [1175.25, 1176.03]$~\AA), and \ion{C}{3} wavelength-integrated line flux ($\lambda = [1294.12, 1294.98]$~\AA). For the white light flux, we masked the $\lambda < 1080$~\AA\ due to the lower signal-to-noise at these wavelengths. We repeated this process for the entire sample of flares. We note that Flare R (Figure~\ref{fig:visit3}) has an extended decay, which could be due to the presence of lower energy flares we are unable to resolve. This event could be influencing our ability to accurately calculate the true $E$ and $ED$ for Flare S as the flux does not return to the pre-flare baseline before Flare S begins; this can be seen in both the white light and \ion{C}{3} light curves and flare properties. The measured properties of all flares are presented in  Table~\ref{tab:Flares_vis3}.

\section{Searching Flare-Affiliated for Non-Thermal Processes}\label{sec:SearchOZ}

A detailed analysis of the behavior of individual emission lines during flare events can yield greater insight into the overall physical mechanisms driving stellar flares. It is only with high signal-to-noise observations of M~dwarfs during large flares, such as those presented here, can we probe these effects.  

\subsection{The Orrall-Zirker Effect}\label{sec:OZEffect}

The Ly$\alpha$ emission line was masked in the observations to avoid detector saturation from 1206 - 1220 \AA. Even with this mask,  the spectral range is sufficient to search for evidence of this effect in the HST/COS data set. Should a non-thermal proton beam be present during these flares, we would expect to see the red wing emission enhancement during the early impulsive phase of a flare \citep{woodgate1992}, when much of the atmosphere among the flaring areas is neutral. Thus, we calculated spectroscopic light curves of the red wing (1220 - 1233.37 \AA) and blue wing (1197.97 - 1206 \AA) of the Ly$\alpha$ emission line to detect the OZ effect. Within the blue wing, we masked \ion{S}{3} ($\lambda$ = 1200.97 \AA) and \ion{Si}{3} ($\lambda$ = 1206.53, 1207.51 \AA), as identified using the HST/STIS spectrum of AU~Mic presented in \cite{rockcliffe2023}. This left a sufficient baseline to determine if the blue wing responded asymmetrically to the red wing in the observed flares. This analysis was repeated for six of the highest-energy flares in our data set: Flares B, J, K, M, R, and T. From the flare properties calculated in Section~\ref{sec:flares}, we can determine that these 6 high-energy flares have \ion{Si}{3} energy $\geq$ the \ion{Si}{3} energy from the flare identified in \cite{woodgate1992}, where the OZ effect was first tentatively detected.

\begin{figure}
    \centering
    \includegraphics[width=1\linewidth]{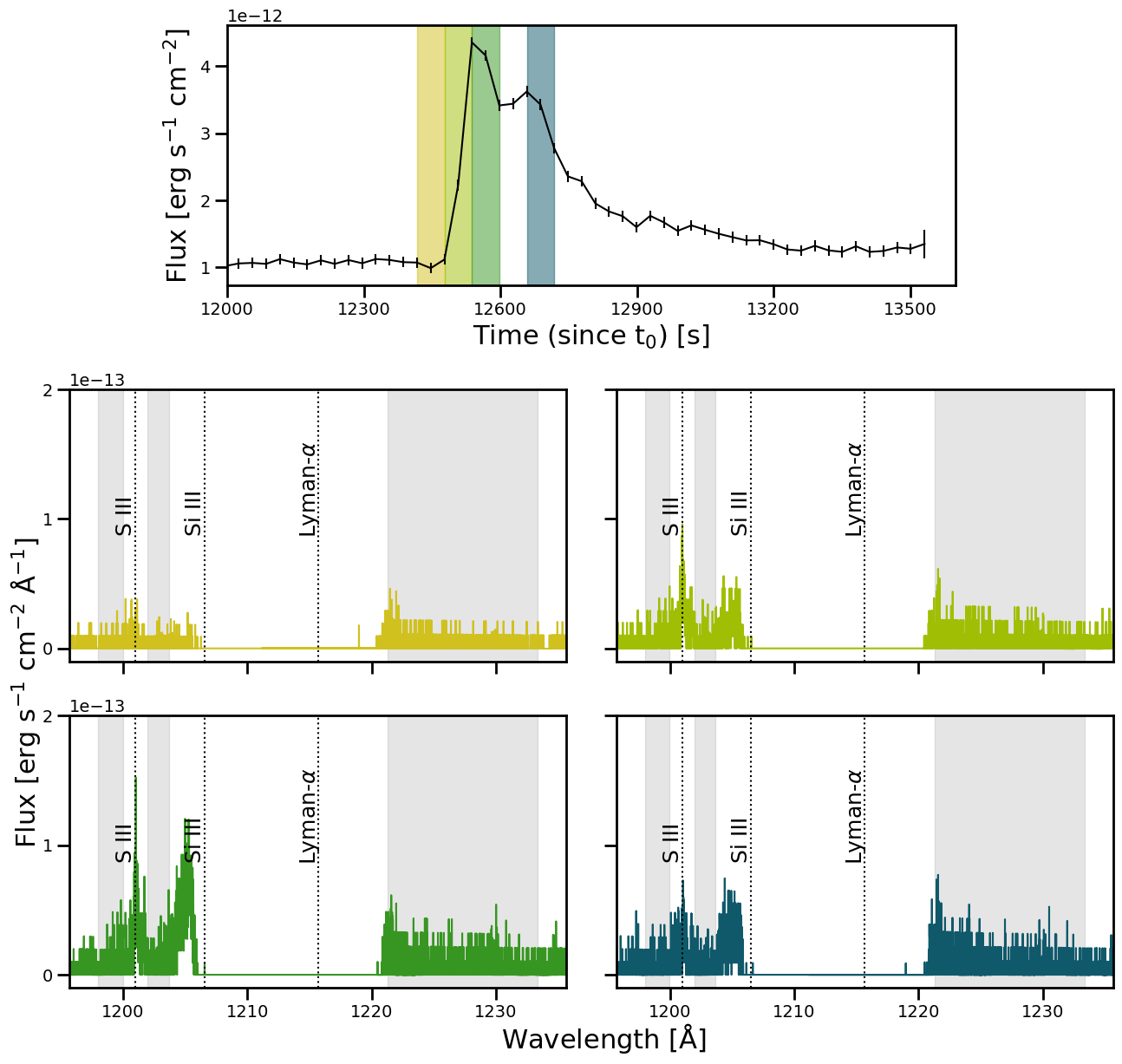}
    \caption{An example of a non-detection of the OZ effect in Flare B in Visit 1. Here, an uneven enhancement of the red wing versus the blue wing of the Ly$\alpha$ emission line (1215.67 \AA) during the impulsive phase of the flare is not observed. The OZ effect predicts an emission enhancement 5.6 - 17.7 \AA\ red-ward of the core of the Ly$\alpha$ emission line compared to the blue wing \citep{kerr23}. The region in which an asymmetric red enhancement is expected is highlighted in gray.  A similar displacement to the blue is indicated in gray, excluding the emission lines. In this flare we saw significant enhancement of the \ion{S}{3} ($\lambda$ = 1200.97 \AA) and \ion{Si}{3} (($\lambda$ = 1206.53 \AA\ or $\lambda$ = 1207.52 \AA) emission lines labeled in this figure. }
    \label{fig:flareBnon}
\end{figure}

We present example time-dependent spectra of the Ly$\alpha$ wings during Flare B in Figure~\ref{fig:flareBnon}. We show the spectra of AU~Mic pre-flare (yellow), the flare impulsive phase (light green), the flare peak (green), and the secondary flare peak (blue). The maximum change in the blue wing flux is consistent with the maximum change in the red wing flux to within $1.5\sigma$. Thus, we find there is no strong evidence of the OZ effect during this flare event. We repeated this analysis for Flares J, K, M, R, and T, which are all consistent with the findings for Flare B.

\subsection{Observed Carbon Blue Enhancement}\label{sec:blue_en}

We describe a blue enhancement in Flare B in two emission lines: \ion{C}{3} ($\lambda_{\text{cen}} = 1175.64$ \AA) and \ion{C}{2} ($\lambda_{\text{cen}} = 1335.095$ \AA).   The centroids of single Gaussian fits are considered as the overall blueshift for the lines. For Flare B, we calculated a maximum blueshift of $333.42^{+120.56}_{-98.34}$~km~s$^{-1}$ in \ion{C}{3} and $200.97^{+126.73}_{-126.84}$~km~s$^{-1}$ in \ion{C}{2}.  Despite the lines being complicated mixtures of several transitions, a single Gaussian fit to each resolved component is sufficient to elucidate general velocities.

\begin{figure}
    \centering
    \includegraphics[width=1\linewidth]{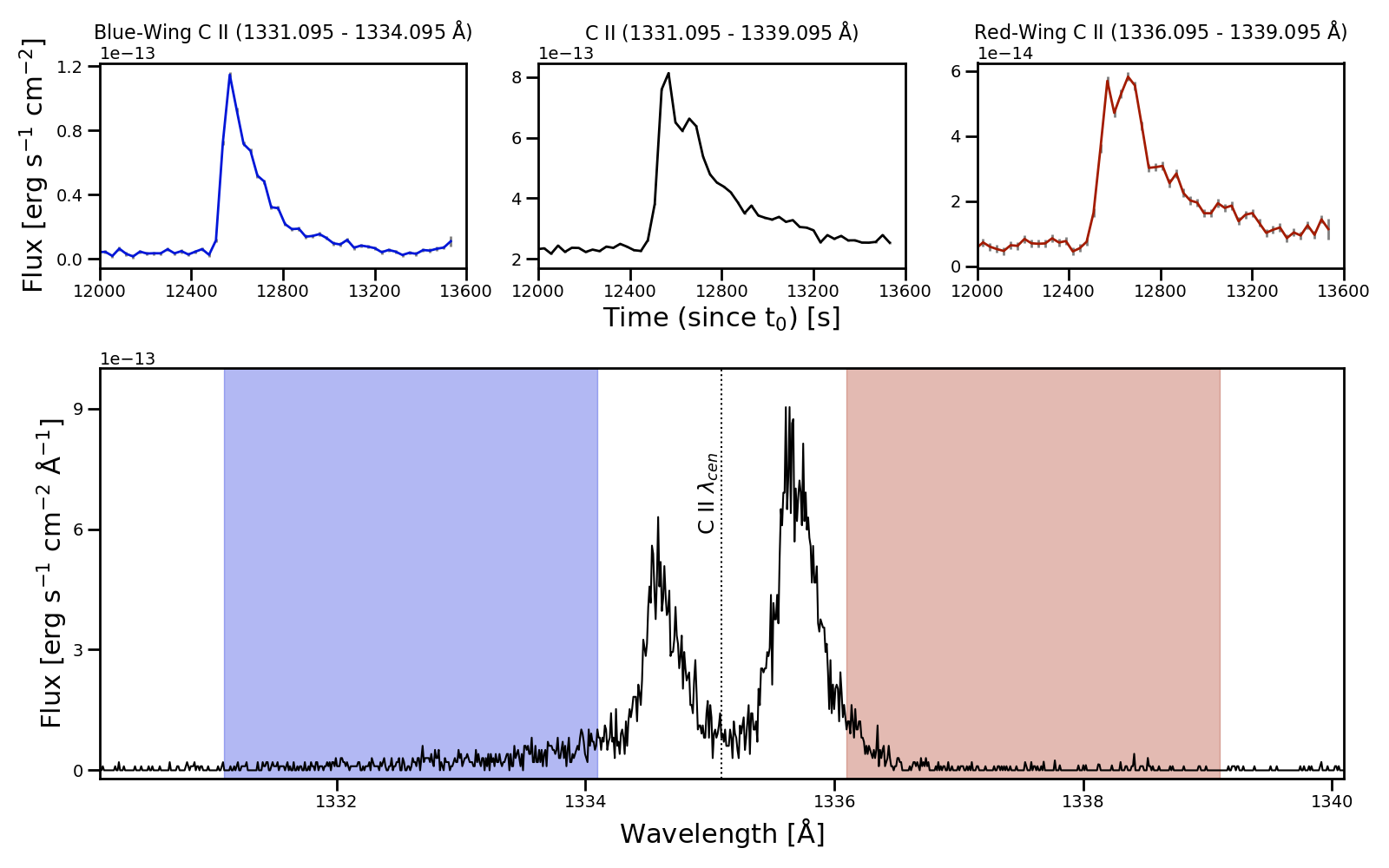}
    \caption{Light curves of Flare B for the blue wing (blue; top left), the entire wavelength (wavelength from start of blue wing to end of red wing; black; top middle), and the red wing (red; top right) of the \ion{C}{2} emission line ($\lambda_{\text{cen}} = 1335.095$ \AA). We highlight the different responses based on the different components of the emission line. In particular, we find that the blue wing has a stronger response to the first peak in Flare B, whereas the red wing has a stronger response to the second peak. This highlights the unique emission line responses that occur during individual flare events.}
    \label{fig:fig6}
\end{figure}

\begin{figure}
    \centering
    \includegraphics[width=1\linewidth]{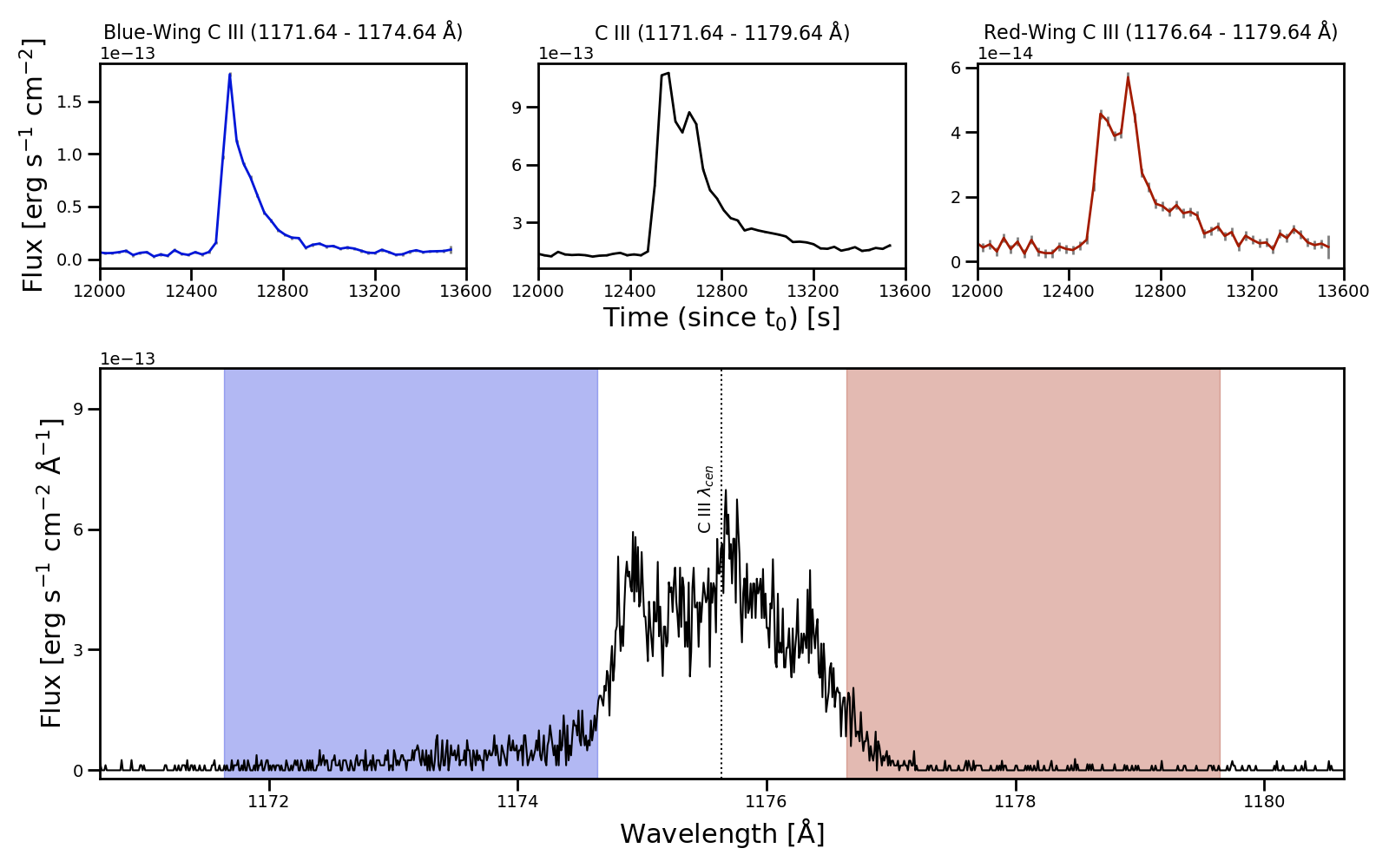}
    \caption{Same as Figure \ref{fig:fig6} but for the \ion{C}{3} emission line. We see the same behavior in \ion{C}{3} as we do for \ion{C}{2}. In particular, note the strong blue asymmetry in the spectrum, whereas the red wing predominantly responds in the second flare peak.}
    \label{fig:fig7}
\end{figure}

Flare B is a large triple-peaked flare event and the longest-lived event observed in the FUV. We present the blue wing, core, and red wing light curves of \ion{C}{2} and \ion{C}{3} in Figures \ref{fig:fig6} and \ref{fig:fig7}, respectively. Both figures show that the blue wings of \ion{C}{2} and \ion{C}{3} appear to respond mainly to the first peak in Flare B, which is further evidence that the second peak in Flare B is a second flare due to its different spectral responses. In the second peak, the blue enhancement of the \ion{C}{3} and \ion{C}{2} spectral lines is not as pronounced as in the first peak. The line response in the second peak seems to be similar to other lines during the flare. In fact, we see in \ion{C}{2} and \ion{C}{3} that the second flare in the Flare B event has a stronger response in the red wing than the first flare. This highlights the unique emission line response for sequential flares. 

We searched for evidence of blue wing enhancements across all other emission lines in our observations. For emission lines with sufficient signal-to-noise (e.g., \ion{N}{5}, Ly$\alpha$ wings), we were unable to detect any asymmetrical responses between the blue and red wings of those lines. All other emission lines lacked the signal-to-noise to sufficiently model the line profile.

\subsection{Modeling Chromospheric Evaporation}\label{sec:modelchromo}

If the blue-enhancement seen in \ion{C}{2} and \ion{C}{3} is due to chromospheric evaporation, this would imply that the gas emitted at $10^{4.2 - 4.8}$~K is being accelerated upwards through the flare loop. The blueshift would be seen at the foot points of the soft X-ray (SXR) loop depending on its temperature. Here, we aim to determine if models of solar-like chromospheric evaporation can be used to explain the observed \ion{C}{2} blue enhancement.

We used a 1D radiative-hydrodynamic M-dwarf stellar flare model from the grid described in \cite{kowalski2024} to model the spectrum of \ion{C}{2} during Flare B. We generated a flare atmosphere model with the following assumptions: (i) peak injected electron beam energy flux density of $2\times 10^{12}$ [erg cm$^{-2}$ s$^{-1}$], (ii) a low energy cutoff of 37 keV, and (iii) a power law index of $\delta$ = 2.5. We chose this model because it is able to reproduce M dwarf optical hydrogen lines well when combined with a model for the continuum \citep{2022Kowalski, 2025Kowalski}. The chosen model also produces upflows of several hundred km~s$^{-1}$ that could inform our understanding of mass motions during the flare events. This model is a higher energy extension of the F11 solar flare model in \cite{2005Allred}, which produces prominent blueshifts in cool temperature (T$ \leq 10^{4.7}$~K) flare lines.

We modeled individual time steps between the impulsive and decay phases of Flare B using the RH radiative transfer code from \cite{2001Uitenbroek}. We evaluated the model for Flare B between t$=0-9$~seconds at $\Delta$ t $= 0.2$~second time-intervals. Note that the model timescale is different than the timescale of Flare B, which lasts $1166.49 \pm 28.23$ seconds (\ref{tab:Flares_vis3}). The models in the grid  represent a superposition of short bursts; nonetheless analysis of individual time steps is necessary to understand how the atmosphere responds in a spatially unresolved observation. To synthesize the C II emission line, we use the model atom from \citep{2015RathCarl}.  A handful of time steps required modification to the number of depth points ($N_\textrm{dep}$) included in the model. Initially, the model had $N_\textrm{dep} = 250$. We had to reduce $N_\textrm{dep}$ to 180 in most cases,  excluding heights in the upper corona, to achieve convergence. As the populations of C II are very small in these hot regions, truncating the atmosphere in this way does not affect our results.  Additionally, time steps from $t = 2 - 3.2$~s required $N_\textrm{dep} < 150$. We concluded this would be an insufficient height representation for an accurate stellar flare model. Therefore, we do not include models at $t = 2 - 3.2$~s for \ion{C}{2} in our analysis. All models are calculated at 5 emergent rays ($\theta = $ 18.19$^{\circ}$, 38.74$^{\circ}$, 60.00$^{\circ}$, 76.70$^{\circ}$, and 87.13$^{\circ}$), and are summed using Gaussian quadrature to give an emergent radiative flux spectrum.

\begin{figure}
    \centering
    \includegraphics[width=0.45\textwidth]{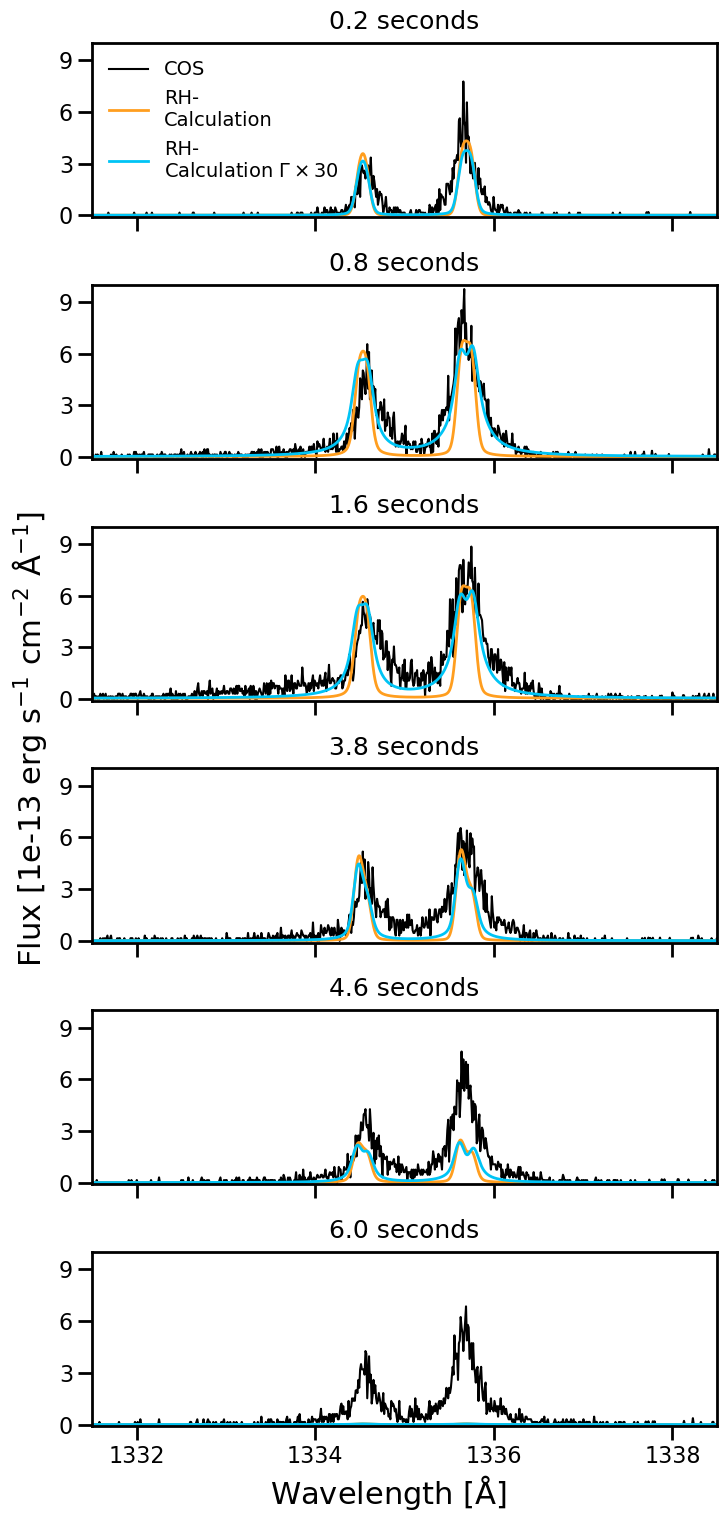}
    \caption{The observed \ion{C}{2} spectral lines during Flare B at different times throughout the flare \citep{2015RathCarl, 2001Uitenbroek}.  Models from the RH code are shown in orange (without damping factor) and blue ($\Gamma \times 30$).   The model time step 6.0 seconds appears nearly flat here. However, the spectral line shape is still present, but the flux at 6.0 seconds is only $\sim 5.3 \%$ of the flux at 4.6 seconds. The model surface flux has been scaled to the observations and convolved to the spectral resolution of the COS/HST observations. The models are scaled by optimizing the projected solid angle of AU Mic on the sky to the observations, each model is scaled by the same factor across time.}
    \label{fig:CIImodel}
\end{figure}

The model comparisons are shown in Figure~\ref{fig:CIImodel}. While every model has limitations, we can draw several conclusions concerning our observation of the response of the \ion{C}{2} emission lines during Flare B from the \cite{kowalski2024} model.  First, we find that the indirect emerging ray ($\theta$ = 87.13$^{\circ}$) contributes to line broadening, while the direct emerging ray ($\theta$ = 18.19$^{\circ}$) contributes to the spectral line strength.   However, none of the rays or the flux  account for the discrepancy between the blue and red enhancements of the \ion{C}{2} spectral line; there is no significant redshift or blueshift in the model for \ion{C}{2}. This chromospheric evaporation model does not explain the observed blue enhancement of the \ion{C}{2} spectral line, which could be indicative of a lack of chromospheric evaporation during this flare or a deficiency in current models.

The source of symmetric broadening in non-hydrogenic lines in solar flares is yet not well understood \citep{2023kerr}.  We follow \citet{2019Zhu} and multiply the Stark-B\footnote{Sahal-Bréchot, S., Dimitrijević, M.S., Moreau N., 2025. STARK-B database, [online]},\footnote{http://stark-b.obspm.fr [June 9, 2025]. Observatory of Paris, LERMA and Astronomical Observatory of Belgrade} damping factor ($\Gamma$) for the C II lines by a factor of 30 \citep{Mahmoudi:2004, Mahmoudi:2005, Larbi-Terzi:2012}. The model with $\Gamma$ $\times$ 30 for \ion{C}{2} (Figure \ref{fig:fig6} better accounts for the shape of the spectral line throughout the flare. However, the model still does not explain the discrepancy between the blue and red enhancements of the \ion{C}{2} spectral line. Quantifying the badness of fit for the models, we find that the RH calculation $\Gamma$ $\times$ 30 had a reduced ${\chi}^2$, $\tilde{\chi}^2 = 0.87 - 2.89$ throughout all the times modeled in Figure \ref{fig:CIImodel}, and we find that the RH calculation without the added $\Gamma$ factor of 30 had a reduced ${\chi}^2$, $\tilde{\chi}^2 = 1.48 - 6.94$.  For the \ion{C}{2} models without the added $\Gamma$ factor of 30, the general line broadening is not well accounted for, as is apparent in Figure \ref{fig:CIImodel}. There is also a poor representation of the intensity of the two different lines in the \ion{C}{2} doublet, with the bluer \ion{C}{2} line ($\lambda$ = 1334.53 \AA) being less intense than the redder \ion{C}{2} line ($\lambda$ = 1335.66 \AA) in the observations but less so in the model.

\section{Discussion}\label{sec:discussion}

\cite{woodgate1992} remains the only tentative detection of the OZ effect to date. Non-thermal proton beams may still be present in flares on AU Mic; however, the new observations exclude copious fluxes of low-energy protons \citep[$< 1$ MeV;][]{kerr23} that cause the OZ effect. Higher energy protons in solar/stellar flares could potentially be detected with $\gamma$-ray and X-ray observations \citep{2020Kerr, 2025Battaglia}.

We compare the observed \ion{C}{2} blueshifts against a well-established chromospheric evaporation model from \cite{kowalski2024}; we found this model did not account for the observed asymmetries, which has led us to propose a second possible mechanism. Filament eruptions often accompany solar flares; the same could be true here. Filaments form in the lower atmosphere and contain cool, dense plasma suspended by magnetic fields. Destabilization and filament eruptions could occur as a result of the magnetic reconnection event and vice versa. The filament eruption happens in the upper threaded loop during this process \citep{2001Moore}. After the low-lying filament erupts, pulling the flare loop up through the atmosphere, the reconnection event happens underneath \citep{shibata1995}. 

In \cite{ikutash2024}, a one-dimensional free-fall model and a pseudo-2D MHD model were used to infer the direction of the ejection, changes in height, and the possible development of a magnetic loop during a flare on EK Draconis. A significant redshift of the H $\alpha$ spectral line was predicted; this magnitude of redshift is not present in the \ion{C}{2} or \ion{C}{3} spectral lines during Flare B. We speculate that the blue enhancement observed here for \ion{C}{3} and \ion{C}{2} could be explained by emission coming from the expansion of the legs of the filament. Solar IRIS observations reveal that \ion{C}{2} and \ion{Mg}{2} exhibit blue shifts attributed to chromospheric evaporation \cite{Li:2019}. Additional analyses of IRIS observations of these spectral lines may provide further insights into spectral tracers of filament eruptions (e.g., \cite{Panos:2018}). This expansion would produce an initial blueshift, followed by a redshift as the filament expands. The phenomenon was reported for the H $\alpha$ line during a superflare on EK Draconis, in \cite{ikutash2024} the H $\alpha $ line variance was modeled during a flare, and it was found that filament eruptions could explain the variance observed. The blue enhancement caused by the filament eruptions has a particular velocity profile starting at velocities of a few tens of km/s up to hundreds of km/s back to zero, and then there is a redshift later on; this process is illustrated in Figure~10 of \cite{ikutash2024}. In \cite{ikutash2024}, there was an initial acceleration of over 500 km/s for the H $\alpha$ spectral line, decelerating to -200 km/s. During Flare B, there was an initial similar blue acceleration of \ion{C}{2} and \ion{C}{3}, in the range of a few hundred km/s. 

Similar models to those used here have been employed for solar flares, and similar line broadening modeling deficiencies have been identified in other spectral lines, such as the Mg II resonance lines in the near UV \citep{2019Zhu, Hawley:2007}. Current abilities to model broadening of \ion{Mg}{2} for solar flares fall drastically short \citep{2019Zhu}.  Our research highlights that a similar problem potentially exists for \ion{C}{2}. \cite{2019Zhu} introduced a Stark width factor of 30. We explore the role of this for \ion{C}{2}, shown in Figure~\ref{fig:CIImodel}. By multiplying $\Gamma$ by a factor of 30 for  the \ion{C}{2} resonance lines, the model is better able to describe the broadening throughout the flare; this suggests that an unknown source of atomic line broadening may also explain the \ion{C}{2} widths in stellar flares.  The enhanced symmetric broadening extends far into the wings (Figure \ref{fig:CIImodel}, t$=0.8 - 3.8$~s).  Determining the physical origin of the missing amounts of the broadening would thus improve the interpretation of asymmetric wings due to mass motions.  In future work, we plan to explore the role of this parameter in more detail while including the \ion{C}{3} lines (which are not included in the \citealt{2015RathCarl} model atom) in the calculations.  

\begin{figure}
    \centering
    \includegraphics[width=1\linewidth]{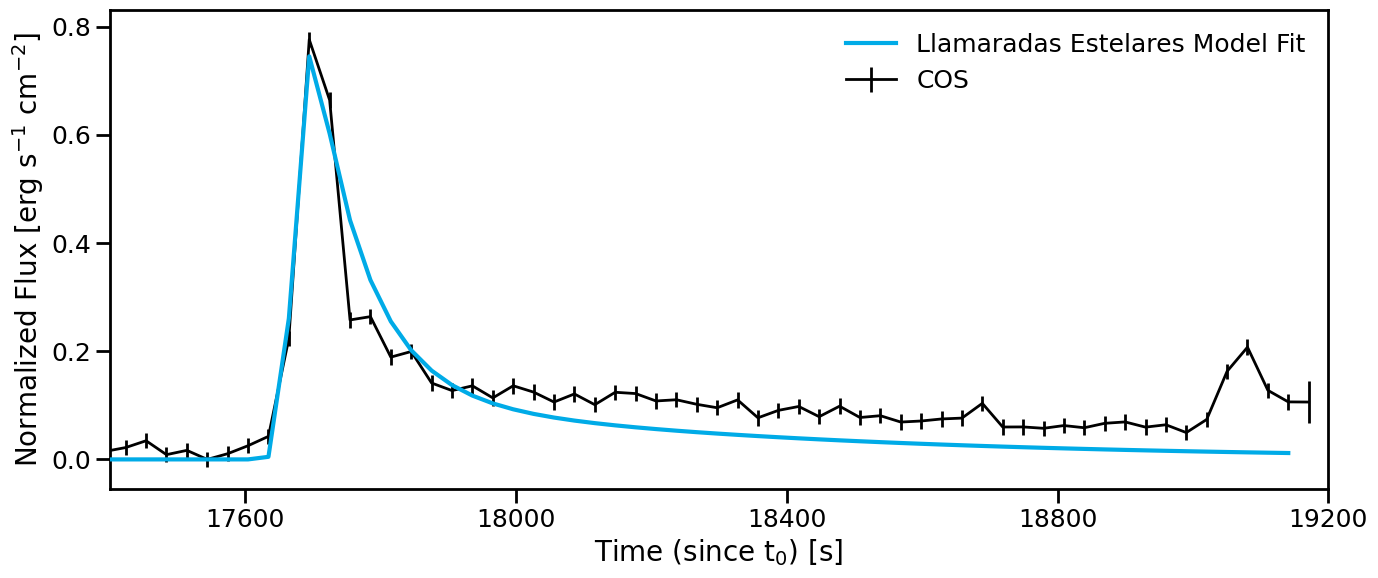}
    \caption{An example of a model fit to Flare R with the Llamaradas Estelares \citep{tovar22} analytical flare model, Flare R is the highest energy flare in Visit 3 with E$_\textrm{flare} = (1.292 \pm 0.176) \times 10^{31}$\ erg. The extended enhanced tail of Flare R is not accurately captured by the model fit; the model fit underestimates the flux during the decay phase of this flare.}
    \label{fig:FlareR_model}
\end{figure}

The flare energies and equivalent durations presented here were derived from the light curve. Other work have reported these values derived from a best-fit model. We attempted to model these FUV flares with the open-source code Llamaradas Estelares \citep{tovar22}, an analytical flare model that was optimized for optical flares. We present an example fit to Flare R in Figure \ref{fig:FlareR_model}. The Llamaradas Estelares flare model was unable to capture the decay phases of several flares, including Flare R, accurately. In particular, the model was unable to account for the extended tail of Flare R, which could be due to subsequent sympathetic flares or persistent emission lines during the decay phase. We emphasize exercising caution in future work when using this model to calculate flare energy (E) and duration, as it has not consistently provided an accurate fit in this set of flares. A more flexible model is needed for modeling FUV flares.  

\section{Conclusions and Future Work}\label{sec:conclusion}

In this paper, we present a reanalysis of \textit{Hubble Space Telescope}/Cosmic Origins Spectrograph FUV observations of the 25~Myr planet-host AU Mic. Our analysis focused on searching for evidence of non-thermal processes across the 20 identified flares in these observations. We conclude the following.

\begin{itemize}
    
    \item We find no evidence of the OZ effect in the high-energy flares analyzed (Section~\ref{sec:OZEffect}); \cite{woodgate1992} remains the only tentative detection of the OZ effect on another star to-date. Our non-detections provide insight into the frequency of proton beams in M-dwarf flares and can be used to enhance our models of such events, which currently lack energy transport from non-thermal protons.

    \item We find evidence of blue wing enhancements in two spectral lines, \ion{C}{2} ($\lambda_{\text{cen}} = 1335.095$ \AA) and \ion{C}{3} ($\lambda_{\text{cen}} = 1175.64$ \AA), during Flare B (Section~\ref{sec:blue_en}). We propose these enhancements may be caused by filament eruptions or chromospheric evaporation. However, we note that the observations may favor filament eruptions, as we tested the observed blueshifts against the well-established chromospheric evaporation model presented in \cite{kowalski2024}. 
    
    \item We modeled the line strength and shape of \ion{C}{2} using the electron beam heating grid provided by \cite{kowalski2024} and found that under the assumptions of the current grid, we are unable to accurately recreate the spectral line strength and broadening of \ion{C}{2} (Section~\ref{sec:modelchromo}). We find this broadening is better explained by multiplying the model by the Stark-B damping factor, $\Gamma = 30$.
    
    \item Our work highlights the lack of understanding of the response of specific emission lines during M dwarf flares. The results of this study will improve our modeling assumptions of M dwarf flares and may have implications for modeling exoplanet atmospheric responses to stellar flares.

\end{itemize}

Additional archival and future observations of AU Mic are important for understanding the occurrence of the OZ effect (e.g., HST GO 17613). AU~Mic was observed with HST/STIS observations in 2020-2021 (HST-GO-15836; PI: Newton). This instrument configuration is centered on Ly$\alpha$ \citep{2023Rockcliffe}. These observations could potentially show the presence of the OZ effect and should be analyzed in future work. 

Here, we focused on using a well-established chromospheric evaporation model from \cite{kowalski2024} to model the \ion{C}{2} line. The work presented here could be extended to modeling \ion{C}{3}. Additional insights from other carbon emission lines could help solidify whether the blue wing enhancements seen during flare events are driven by chromospheric evaporation or an alternative mechanism. Additional, simultaneous observations in the soft X-ray and other ultraviolet wavelengths could also be used to better understand this phenomenon.

In solar flares, we expect the transition from blueshifts to redshifts to occur at about 1 MK, according to \cite{Milligan&Dennis}. However, in these stellar flares, we observe that this does not hold. We are observing blueshifts in lines that are significantly cooler than 1 MK, \ion{C}{2} \citep[log$_{10(}T_\textrm{form} = 4.20$ K;][]{hawley2003} and \ion{C}{3} \citep[log$_{10(}T_\textrm{form} = 4.80$ K;][]{hawley2003} in Flare B. The chromospheric evaporation scenario that we rule out for the flares of AU Mic is driven by high-energy electron beams. Alternatively, high-energy proton beams (consisting of E $>$ 10 MeV protons and thus not leaving signatures in the red wing of Ly$\alpha$) could lead to deeper evaporation of cooler material in flares \citep{2024Sadykov}. We plan to investigate this with the RADYN code in future work \citep{test}.

The prolonged enhancement during the decay phase of several flares, as well as complex flare events such as Flare B, suggests that a more detailed flare model is required to explain an observed persistent heating after high-energy events, which a single heating phase cannot explain during the flare. These advancements need to occur both from modeling the morphology of the flare light curve and in the 1-D radiative-hydrodynamic models. For example, it could be the case that the broadened \ion{C}{2} spectral lines during Flare B (Figure~\ref{fig:CIImodel}) may be due to the complex nature of this multi-flare event.

The results of this work are important for understanding flare physics on M dwarfs and for understanding the impact of stellar activity on short-period planets. We require a comprehensive understanding of stellar flare atmospheric physics to be able to model and address whether rocky planets orbiting M dwarfs could retain their atmospheres and be habitable, or not. A thorough understanding of the physics of accelerated non-thermal protons and electrons in energy transport is crucial to this endeavor.

\vspace{1em}
\noindent\textbf{Acknowledgments.}
ADF acknowledges funding from NASA through the NASA Hubble Fellowship
grant HST-HF2-51530.001-A awarded by STScI.

The HST data presented in this article was obtained from the Mikulski Archive for Space Telescopes (MAST) at the Space Telescope Science Institute. The specific observations analyzed can be accessed via \dataset[doi: 10.17909/bh63-4j34]{https://doi.org/10.17909/bh63-4j34}.
\vspace{1em}

\newpage
\appendix

We present the measured energies and equivalent durations in Table~\ref{tab:Flares_vis3}. We present a recalculation of the white-light (WL) flares for Flares A - M. Additionally, we present the measured flare properties of Flares A - M as observed in the \ion{Si}{3} and \ion{C}{3} emission lines. We repeat this for the new flares identified in this work (N - T).

\begin{table*}
    \centering
    \caption{Re-calculated broadband (white-light) light curve flare properties, and newly calculated \ion{Si}{3} ($\lambda_{cen}$ = 1294.55 \AA), and \ion{C}{3} ($\lambda_{cen}$ = 1175.64 \AA) flare properties for visit 1 and visit 2. White light, \ion{Si}{3} ($\lambda_{cen}$ = 1294.55 \AA), and \ion{C}{3} ($\lambda_{cen}$ = 1175.64 \AA) flare properties of flares identified in visit 3. The Equivalent Durations for flares S and T are not included because accurate end times for flares S and T were unable to be determined due to the orbital gaps. Due to the low amplitude of some flares in \ion{Si}{3} and \ion{C}{3} the flare energy (E) and equivalent durations were unable to be properly constrained. t$_{peak}$ is seconds after the start time of the exposure. }
    \label{tab:Flares_vis3}
    \begin{tabular}{|c|c|c|c|c|c|c|c|c|c|}
       \hline
       Flare  & t$_{peak}$ & E$_\textrm{WL}$  & ED$_\textrm{WL}$& E$_\textrm{Si III}$& ED$_\textrm{Si III}$ & E$_\textrm{C III}$  & ED$_\textrm{C III}$  \\
         & [s] & [$10^{29}$ erg] &  [s] & [$10^{27}$ erg] & [s] & [$10^{29}$ erg] & [s] \\
       \hline
       \hline
       A  & 6388 & $14.36 \pm 5.76$ & $136.74 \pm 0.773$ & --- & --- & $0.35 \pm 0.079$ & $42.90 \pm 10.19$\\
       B  & 12531 & $122.5 \pm 14.15$ & $1166.49 \pm 28.23$ & $56.84 \pm 11.05$ & $1538.56 \pm 42.85$ & $10.94 \pm 0.21$ & $1328.01 \pm 254.63$\\
       C  & 16935 & $15.99 \pm 4.27$ & $152.24 \pm 0.70$ & $3.61 \pm 3.31$ & $97.66 \pm 2.80$ & $0.51 \pm 0.060$ & $61.49 \pm 14.45$\\
       D  & 17985 & $12.37 \pm 2.69$ & $117.80 \pm 0.42$ & $12.56 \pm 2.13$ & $339.98 \pm 10.16$ & $1.38 \pm 0.038$ & $167.56 \pm 37.87$\\
       E  & 23049 & $5.83 \pm 3.46$ & $55.48 \pm 0.36$ & --- & --- & $0.16 \pm 0.047$ & $18.98 \pm 4.24$\\
       F  & 24819 & $6.93 \pm 3.76$ & $65.98 \pm 25.80$ & $1.70 \pm 2.70$ & $46.14 \pm 1.42$ & $0.13 \pm 0.047$ & $15.23 \pm 3.58$\\
       G  & 1740 & $3.45 \pm 2.089$ & $31.051 \pm 28.32$ & $4.47 \pm 1.64$ & $264.65 \pm 32.50$ & $0.26 \pm 0.029$ & $32.65 \pm 31.55$\\
       H  & 17515 & $5.078 \pm 3.80$ & $45.68 \pm 0.27$ & --- & --- & --- & --- \\
       I  & 22993 & $2.72 \pm 1.52$ & $24.46 \pm 0.18$ & --- & --- & $0.29 \pm 0.021$ & $37.29 \pm 8.65$\\
       J  & 23473 & $11.13 \pm 2.28$ & $100.12 \pm 0.29$ & $10.61 \pm 1.81$ & $628.79 \pm 8.96$ & $1.44 \pm 0.034$ & $183.58 \pm 39.29$\\
       K  & 23653 & $23.20 \pm 3.80$ & $208.66 \pm 0.58$ & $27.77 \pm 3.046$ & $1646.011 \pm 22.93$ & $2.69 \pm 0.058$ & $343.40 \pm 70.42$\\
       L  & 23983 & $11.23 \pm 2.66$ & $101.025 \pm 0.40$ & $10.65 \pm 2.11$ & $631.32 \pm 8.94$ & $0.82 \pm 0.039$ & $104.39 \pm 23.42$\\
       M  & 24493 & $24.29 \pm 4.56$ & $218.46 \pm 0.72$ & $17.55 \pm 3.64$ & 1$040.16 \pm 14.058$ & $1.71 \pm 0.069$ & $217.79 \pm 44.76$\\
       N  & 6090 & $8.74 \pm 3.06$ & $66.46 \pm 0.23$ & $8.21 \pm 2.46$ & $359.23 \pm 7.056$ & $0.86 \pm 0.044$ & $100.59 \pm 24.071$\\
       O  & 11622 & $9.82 \pm 4.59$ & $74.67 \pm 0.34$ & $4.45 \pm 3.67$ & $194.29 \pm 3.87$ & $0.63 \pm 0.065$ & $73.24 \pm 18.30$\\
       P  & 12616 & $3.36 \pm 1.53$ & $25.52 \pm 0.083$ & $3.77 \pm 1.22$ & $164.79 \pm 3.28$ & $0.27 \pm 0.022$ & $31.57 \pm 7.60$\\
       Q  & 16941 & $5.89 \pm 1.53$ & $44.76 \pm 0.23$ & $6.55 \pm 1.24$ & $286.44 \pm 5.67$ & $0.41 \pm 0.022$ & $47.74 \pm 11.42$\\
       R  & 17694 & $129.2 \pm 17.61$ & $982.039 \pm 2.80$ & $72.57 \pm 14.76$ & $3173.81 \pm 53.16$ & $12.24 \pm 0.27$ & $1424.82 \pm 294.75$\\
       S  & 19081 & $14.60 \pm 2.20$ & --- & $11.15 \pm 1.78$ & --- & $1.70 \pm 0.030$ & ---\\
       T  & 24795 & $25.18 \pm 1.81$ & --- & $21.22 \pm 1.50$ & --- & $1.72 \pm 0.027s$ & ---\\
       \hline
    \end{tabular}
\end{table*}

\bibliography{sample631}{}
\bibliographystyle{aasjournal}

\end{document}